\documentstyle[sprocl]{article}
\input{psfig}

\def\be{\begin{equation}}
\def\ee{\end{equation}}
\def\bea{\begin{eqnarray}}
\def\eea{\end{eqnarray}}

\def\gtap{\raisebox{-.55ex}{\rlap{$\sim$}} \raisebox{.4ex}{$>$}}
\def\gsim{\mathrel{\gtap}}

\begin{document}
\title{A STUDY OF THE EFFECTS OF PREHEATING\footnote{Talk given at 
18th {\it Texas Symposium on Relativistic Astrophysics}, Chicago, IL, 
15-20 Dec 1996.} }
\author{I. TKACHEV }
\address{Department of Physics, The Ohio State University, Columbus, OH 43210\\
and\\
Institute for Nuclear Research of the Academy of Sciences of Russia\\
Moscow 117312, Russia}
\maketitle
\abstracts{We review results of a numerical study of 
resonant inflaton decay in a wide range of realistic models and parameters.
Wide enough parametric resonance can withstand the 
expansion of the Universe, though account for the expansion is very 
important for determining precisely how wide it should be.
For example, the effective production of particles with mass ten 
times that of the inflaton requires very large values of the resonance
parameter $q$, $q > 10^8$. For these large $q$, the maximal size of
produced fluctuations is significantly suppressed by back reaction,
$\langle X^2 \rangle \sim 10^{-10} M_{\rm Pl}^2$.
We discuss some physical implications of our results.}

Recently the scenario of inflaton decay and reheating of the Universe 
after inflation has been revised considerably, starting from
the observation~\cite{KLS} that this decay can naturally be in a regime of
a wide parametric resonance. Wide resonance can be effective 
despite the expansion of the Universe. During this explosive process 
(called preheating~\cite{KLS}) particles heavier than inflaton can be created.
This opens possibilities for many interesting and important effects to occur,
including non-equilibrium phase transitions~\cite{effects} and 
generation~\cite{moref} of baryon asymmetry  in old GUT frameworks 
(provided GUT is B-L non-conserving).

Particle content of the theory assumed to be as follows.
First, there is an inflaton field which we denote as $\phi$.
Second, there are products of inflaton decay. We denote them as $X$ and
the mass of corresponding quanta is $m_X$. 
The possibility of the inflaton decay into $X$-quanta assumes that there
is interaction of, say, the form $g^2X^2\phi^2/2$. In reality, there can
be many channels for the inflaton decay and the final answer is the sum over
all species. Third, the model allows for the possibility of symmetry 
breaking with an order parameter $\Phi$. The $X$-particles couple 
to the order parameter, so their mass depends upon it, 
$m_X = m_X(\Phi)$. 
The part of the total potential (we break it into parts as $V=V_1+V_2+ \dots$) 
relevant for the symmetry breaking is 
$V_1 = -\frac{1}{2}\mu^2 \Phi^2 + \frac{1}{2} \alpha X^2 \Phi^2$.

If $X$-particles are abundantly created, the effective mass of $\Phi$-filed
can be written as $m_{\rm eff}^2 = -\mu^2 + \alpha \langle X^2\rangle$. 
The symmetry is restored if $m_{\rm eff}^2 > 0$, or 
$\langle X^2\rangle > \mu^2/\alpha$. In thermal
equilibrium with temperature $T$ one would have $\langle X^2\rangle =T^2/12$. 
The main observation~\cite{effects} is that right after the inflaton decay but long
before equilibrium is established, $\langle X^2\rangle$ is anomalously 
large. The strength of symmetry restoration in this non-equilibrium state 
can exceed that one in thermal equilibrium by many orders of magnitude. 

Let us illustrate this point neglecting the expansion of 
the Universe. Energy density conserves in this case and energy density stored 
in $X$-filed after decay is equal to energy density in initial 
inflaton oscillations, $\rho \sim m_i^2 M^2_{\rm Pl}$. 
Typical energy of $X$-quanta is of 
order of the inflaton mass, $E \sim m_i$. We find 
$\langle X^2\rangle \sim \rho/E^2 \sim M^2_{\rm Pl}$, 
i.e typical scale for the strength of symmetry restoration
is Plankian. This estimate is encouraging but oversimplified. 
First, the expansion is 
very important: while parametric resonance develops, fields are 
red-shifted despite the explosive character of the process. 
Second, the inflaton not always decays completely during the resonance
stage, and due to various back-reaction effects the fast decay can stop at 
much smaller values of $\langle X^2 \rangle$. 
The resulting picture is very model dependent,
and due to complexity requires numerical study, model by model. 

Number density of created particles can be easily determined if 
$\langle X^2 \rangle$ is known. For massive X, which are created 
mostly non-relativistic, we have $n_X=\langle X^2 \rangle /m_X$. Suppose 
the baryon number and CP are violated in decays of $X$ particles (i.e. X are 
GUT leptoquarks) and they are created in sufficient number during parametric 
resonance. This opens a possibility of
BAU generation at preheating~\cite{moref}$^,$\cite{effects}.

Clearly, the  quantity of interest is the maximum value
of $\langle X^2 \rangle$ which can be achieved in inflaton decay. Here I 
review some results for different models.

1. The simplest one is pure $\lambda \phi^4$ model where the inflaton decays
due to self-interaction. Fully non-linear calculation of the decay,
which includes all rescattering and back-reaction processes~\cite{us}, gives
$\langle (\phi-\phi_0)^2 \rangle_{\rm max} = 10^{-7} M_{\rm Pl}^2$.

2. Massless inflaton decaying into $X$-particles, 
$V_2=\frac{1}{4}\lambda \phi^4 + \frac{1}{2}
g^2 X^2\phi^2$. Decay is very inefficient if $X$ are massive~\cite{wide}. 
In a conformally invariant case, typically~\cite{wfnl} $\langle X^2 \rangle
_{\rm max}\approx 10^{-7} 
({100}/{q})^{3/2} M_{\rm Pl}^2$ for $q \gsim 100$, where 
$q \equiv g^2/4\lambda$.
Rescattering effects suppress 
$\langle X^2 \rangle$ as an inverse power of 
q~\cite{wide}$^,$\cite{wfnl}$^,$\cite{PR}.

3. The most complicated and efficient is the case with massive inflaton,
$V_2=\frac{1}{2}m_i^2 \phi^2 +\frac{1}{2}g^2 X^2\phi^2$. Inflaton can decay 
even if $X$ is
considerably heavier than inflaton itself. A good idea of how 
$\langle X^2 \rangle_{\rm max}$ depends upon mass 
$m_\chi \equiv m_X/m_i$ and $q$ (where now $q \equiv g^2 \phi^2(0)/4m_i^2$)
in an expanding universe can be obtained in the Hartree approximation.
It is plotted in Fig. 1 by solid lines~\cite{wide}. We see that to have 
resonance at all in an expanding universe,
even for massless X-particles, one needs $q \gsim 10^{3}$. Required $q$
rapidly grow with $m_\chi$. E.g., to create $X$-particles sufficiently heavy 
to be promising for baryogenesis, $m_\chi\approx 10$, one already needs 
$q \sim 10^8$.
For creation of massless particles expansion becomes unimportant and 
$\langle X^2 \rangle_{\rm max}$ becomes saturated by back-reaction if 
$q \gsim 10^4$. The
Hartree approximation grossly overestimates $\langle X^2 \rangle_{\rm max}$
in this regime (however, this
approximation can be good for X being a multicomponent field, such as a
large multiplet  of $O(N)$ model). 
Results of fully non-linear lattice calculations~\cite{wfnl}, which 
include all effects but are more expensive, are presented in Fig. 1 by 
stars. Data points are presented for massless $X$ at the moment 
when zero mode of inflaton oscillations completely decayed.
Extrapolating these results to $q \approx 10^8$ we anticipate~\cite{wfnl}
$\langle X^2 \rangle_{\rm max} \approx 10^{-10} M_{\rm Pl}^2$.

Fig. 1 is a starting point for calculations of BAU and symmetry restoration
effects at and after preheating.

\begin{figure} 
\psfig{figure=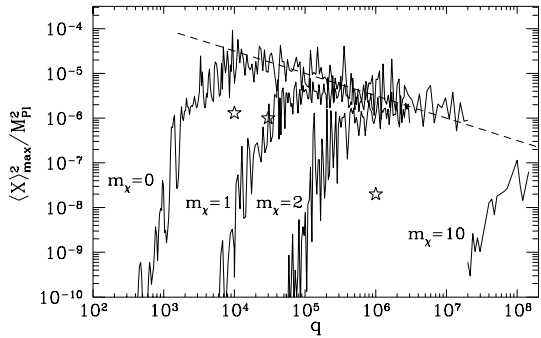,height=1.9in} 
\caption{Variance $\langle X^2 \rangle$ as a function of $q$. 
\label{fig1}}
\end{figure}

\section*{Acknowledgments} 
This work  was supported by DOE Grant DE-AC02-76ER01545 at Ohio State.

\end{document}